# PREPRINT NOTES

## Title of the article:

A Deployment Model to Extend Ethically Aligned AI Implementation Method ECCOLA

## Authors:

Jani Antikainen, Mamia Agbese, Hanna-Kaisa Alanen, Erika Halme, Hannakaisa Isomäki, Marianna Jantunen, Kai-Kristian Kemell, Rebekah Rousi, Heidi Vainio-Pekka, Ville Vakkuri

## Notes:

- This is the authors' version of the work
- Paper has been accepted as a workshop long paper in 2021 IEEE 29th International Requirements Engineering Conference and will be published in the conference proceedings

# A Deployment Model to Extend Ethically Aligned AI Implementation Method ECCOLA

Jani Antikainen
*Faculty of Information Technology*
*University of Jyväskylä*
Jyväskylä, Finland
0000-0003-3367-0492

Mamia Agbese
*Faculty of Information Technology*
*University of Jyväskylä*
Jyväskylä, Finland
mamia.o.agbese@jyu.fi

Hanna-Kaisa Alanen
*Faculty of Information Technology*
*University of Jyväskylä*
Jyväskylä, Finland
0000-0002-8797-3432

Erika Halme
*Faculty of Information Technology*
*University of Jyväskylä*
Jyväskylä, Finland
0000-0003-0750-1580

Hannakaisa Isomäki
*Faculty of Information Technology*
*University of Jyväskylä*
Jyväskylä, Finland
0000-0002-6021-3118

Marianna Jantunen
*Faculty of Information Technology*
*University of Jyväskylä*
Jyväskylä, Finland
0000-0002-8991-150X

Kai-Kristian Kemell
*Faculty of Information Technology*
*University of Jyväskylä*
Jyväskylä, Finland
0000-0002-0225-4560

Rebekah Rousi
*Faculty of Information Technology*
*University of Jyväskylä*
Jyväskylä, Finland
0000-0001-5771-3528

Heidi Vainio-Pekka
*Faculty of Information Technology*
*University of Jyväskylä*
Jyväskylä, Finland
0000-0002-9736-3400

Ville Vakkuri
*Faculty of Information Technology*
*University of Jyväskylä*
Jyväskylä, Finland
0000-0002-1550-1110

*Abstract* — *There is a struggle in Artificial intelligence (AI) ethics to gain ground in actionable methods and models to be utilized by practitioners while developing and implementing ethically sound AI systems. AI ethics is a vague concept without a consensus of definition or theoretical grounding and bearing little connection to practice. Practice involving primarily technical tasks like software development is not aptly equipped to process and decide upon ethical considerations. Efforts to create tools and guidelines to help people working with AI development have been concentrating almost solely on the technical aspects of AI. A few exceptions do apply, such as the ECCOLA method for creating ethically aligned AI -systems. ECCOLA has proven results in terms of increased ethical considerations in AI systems development. Yet, it is a novel innovation, and room for development still exists. This study aims to extend ECCOLA with a deployment model to drive the adoption of ECCOLA, as any method – no matter how good - is of no value without adoption and use. The model includes simple metrics to facilitate the communication of ethical gaps or outcomes of ethical AI development. It offers the opportunity to assess any AI system at any given lifecycle phase, e.g., opening possibilities like analyzing the ethicality of an AI system under acquisition.*

## I. INTRODUCTION

Artificial intelligence (AI) ethics does not yet possess an established definition per se but alludes to the connection between ethics and AI [1]. AI ethics finds itself in a situation in which abstract ethical principles [2, 3] are the focus of research. Simultaneously, practitioners require tangible models and tools on how to turn those abstract principles into actionable work [2, 3]. At the same time, AI is increasingly pervasive in our daily lives as part of the software (SW) products and services we use - many of which have our lives in their hands (e.g., semi-autonomous cars). Cases in which ethical considerations are omitted can manifest into significant physical or cyber-related damage. The realized or potential adverse impacts [4] of such AI -inclusive artefacts have given rise to the recent focus on AI ethics [4]. Yet, even with the increased focus and invested effort by numerous actors from academia [5] and industry, such as Intel [6] and Microsoft [7], the European Commission [8], nation states [e.g., 9] as well as professional technology organizations like IEEE [10], the evolution from a plethora of theory-based principles into actionable practice is proving difficult [11, 12].

Hardly any solutions have been voiced to answer the increasing call for actionable AI ethics models and methods supporting the development of ethically aligned AI systems. Ethically aligned AI development method ECCOLA [3] has emerged to address this challenge and is already in the post-conceptual development phase. It provides evidence showing that "(…) simply the presence of an ethical tool affects ethical consideration, creating more responsibility even in instances where the use of the tool is not intrinsically motivated" [13]. As an actionable tool, ECCOLA facilitates the addressing of AI ethics in an agnostic methodological way suitable for different approaches in software development while promoting Agile methodologies.

The ECCOLA -method instructs its users to create Work Product Sheets (WPS) during every sprint and document WPS justification and further review the actions [3]. Yet, it does not provide instructions or capability to assess given AI SW artefact's overall ethical standing (as a holistic product). The previous holds particularly in terms of the product's ethical coverage at any given lifecycle state or ways to communicate such situations to relevant stakeholders. Stakeholders might represent differing points of view, needs, and concerns about the AI SW artefact, its ethical span, and little if any knowledge of software engineering (SE). Thus, a self-explanatory, rapidly deployable model characterized by ease-of-use is needed to plan, form, and communicate a situational picture of ECCOLA coverage. Such a framework further serves the need to involve all relevant stakeholders of the AI SW artefact (also beyond the artefact's development -phase) as a multi-disciplinary assessment team. A team representing notions and needs that software developers and product owners are not alone equipped to cater for either resource or competence-wise [4].

This paper will present a deployment model for the ECCOLA method, which builds on and extends ECCOLA. The model presented is at the same time a light assessment tool to provide visibility in the overall of the AI artefact's ethical standing. Motivation for the research is driven by the following notions: (1) No tool exists to assess ethically aligned AI SW creation effort (the scope of ECCOLA method), (2) a hypothesis that expands the ECCOLA toolset further incorporates the entire lifecycle of the SW product into its development logic, (3) to identify additional use cases for the ECCOLA method beyond the current use case, (4) to foster the adoption of ECCOLA via proposing an extension to its utilization on areas of planning, prioritization, and validation of ethical considerations, (5) to offer a capability to provide a snapshot of how ECCOLA's ethical considerations are realized at any given point in the AI SW product's lifecycle by embedding the situational picture with easily understandable and conveyable metrics, (6) a need to assure that ECCOLA's adoption and use has the relevant domain expertise involved from potentially very diverse stakeholder landscape, and that the stakeholders are actively involved, (7) addressing transparency, communication, and stakeholder involvement, (8) raising general awareness in AI ethics, (9) and creating a needed dialog between AI ethics theory and practice by researching and developing an artefact which considers theory yet is actionable and practical at the same time.

The authors of this paper agree with Morley et al [2], who argue for the need for the machine learning (ML) community to unite multi-disciplinary researchers into the pro-ethical design of tools and methodologies for ML and AI. This paper answers the call of Morley et al [2] to coordinate effort on applied (ML) ethics by creating more practical tools that are rapidly advanced. Such application expands into production environments and will utilize multi-disciplinary competencies represented by innovators, policymakers, designers, and so on [2]. The study aims to identify gaps while contributing to advances in the first three of six focus areas identified by Morley et al [2]:

*1. the development of a common language.*

*2. the creation of tools that ensure people, as individuals, groups, and societies, are given an equal and meaningful opportunity to participate in the design of algorithmic solutions at each stage of development.*

*3. the evaluation of the tools that are currently in existence so that what works, what can be improved, and what needs to be developed can be identified.*

The design science research (DSR) paradigm [14] is adopted to construct a conceptual model of the envisioned adoption framework for ECCOLA method. Further positioning the paper and the work done within the DSR realm, one should consider the design science research methodology (DSRM) as the chosen research approach on the proposed lines of Peffers et al [15]. Emphasis is on designing and constructing an applicable real-world artefact (the framework), which aims to be generalizable, highly practical, and driven by a strong motivation to solve part of the problem in AI ethics currently experienced between theory and practice.

As the target of the study has been to aim for actionable and practical results, a considerable part of the study has been based especially on the first author's 24 years of ICT and business professional practice, as a reflective practitioner who uses knowledge-in-action [16] as a key tool. Knowledge-in-action refers to research paradigm of pragmatism and aims for both practical and scientific contribution [14].

This paper is aimed for any role involved in AI SW artefact's lifecycle from idea to use who have a need or interest to adopt and deploy AI ethical considerations. The paper is also relevant for researchers interested in AI ethics and the practice of implementing and assessing actionable approaches to the topic. The paper proceeds with Section II elaborating the related work and theories used, followed by presenting the ECCOLA deployment and assessment model in Section III, where Section IV will propose research opportunities and concludes the paper.

## II. ECCOLA APPROACH TO AI ETHICS

This section is divided into two sub-sections. First the authors elaborate on AI ethics while the second section introduces the ECCOLA -method.

### A. AI Ethics

The Introduction Section described how AI ethics do not currently entail an agreed-upon definition, even though the concept is not a new one, dating as far as the 1960s [2]. This paper does not dwell on this debate any further as the problematics quickly add up, for example, by similar terms existing with different definitions, e.g., ethical AI, which [17] defines as "the practice of using AI with good intention."

AI ethics are still primarily macro ethics, principles not grounded into individual decision making or consideration of one's agency and responsibility while solving some real-life problem with software inclusive of AI elements. The latter is echoed by Vakkuri et al [18] as they analyze the current state of AI ethics, finding it as a discussion on guidelines and principles. Unfortunately, according to McNamara et al [19], who studied ACM Code of Ethics principles' impact on practices, came to an outcome of impact been measured either as little or not at all. Analogous findings with McNamara et al are discovered by Vakkuri et al [18] as they summarize literature principles not finding traction on the practice. Perhaps a notion by Jobin et al [20] provides a descriptive take on the state of matters, noting that even these abstractions, the principles, which we found not to connect with practice, are much defined by the number that they appear on numerous AI guideline articles.

As developers are creating (AI) systems, they do so, reflecting on their personal views and values in the design and implementation of such systems. This is thus transferred into the systems themselves [21]. Unfortunately, more challenge for AI ethics is readily available. It has been shown that when the developers encounter ethical questions or issues, they often end up being simplified or entirely neglected [13]. Partly, this is explained by a lack of means to address ethical concerns, i.e., the lack of actionable tools [18]. A more plausible explanation is that ethical issues do not connect with their interests, which heavily hinge upon work concerns [22]. Ultimately, the state of matters is such that developers do not find ethics meaningful [13].

*B. ECCOLA*

The previous section addressed the disconnection of AI ethics principles and practice. Actionable tools (e.g., methods) have been proposed as one potential solution [e.g.,2,3] to bridge the gap. Such tools alone are not the complete answer as they come with further requirements. For example, Abrahamsson and Iivari [23] present a notion that SW developers prefer practical and straightforward methods. Tools, no matter how simple and functional are overlooked. At the same time, developers are allocated the vital role of implementing AI ethics [e.g.,2,3].

Answering the call for practical and actionable approaches to AI ethics, the ECCOLA method [3] has been created and initially validated [24] to have a positive effect on attention to ethical considerations of SW developers and product managers, even when the model is not enforced [13]. ECCOLA is agnostic of the SW development method as well as the context in which it is used. ECCOLA encourages organizations to consider the AI ethics of AI -enabled systems [3]. In the context of Lean software product development [25], ECCOLA positions as methods and practices. It takes advantage of ethical AI principles and concepts that others have created previously.

ECCOLA draws from both the European Commission's [8] and IEEE's [10] ethical AI. ECCOLA is materialized into a set of 21 cards, each representing one element of 8 themes (*analyze, data, transparency, agency & oversight, safety & security, wellbeing, fairness, and accountability*) of AI ethics principles. ECCOLA is meant to be light and to support agile methodologies. Thus its use would be straightforward.

For each iterative development sprint, the cards representing the relevant elements (e.g., *data quality* under *data* -theme) for that sprint are selected, the selection justified and documented. Next, during development tasks the selected cards are reviewed and documented for action taken on them. Last, at the end of the sprint, the planned actions are evaluated, the card deck revised and the cards most relevant for the next sprint are selected.[3]

Each ECCOLA card empowers ethical thinking in the development process via providing motivation for the element it represents (why it is important, why to do), practical ideas on what to do to cover the element and practical example(s) of the element in question.[3]

Although ECCOLA is a rare breed in providing actionable guidance and considerations for otherwise abstract AI ethics principles [3], it is still a relatively novel tool. Hence, it has much potential for development and expansion of the method.

III. THE DEPLOYMENT MODEL

This section will introduce the created deployment model for ECCOLA. The section is split into three sub-sections. The first sub-section will frame the boundaries of what should be considered while developing a deployment model for ECCOLA. The second sub-section describes the created model's process and use. The third sub-section will demonstrate a communicable situation picture snapshot -chart of how different ECCOLA- themes and their elements can be visualized and communicated.

*A. Deployment model considerations*

Among other challenges with AI ethics are the notions that ethics are culturally dependent. It is likely that there will never be consensus on a universal definition of principles [4]. Further, the context in which ethics are applied plays a significant role. Context influences the types of ethical considerations that most matter in a AI system [3,4]. S*oftware developers* and *product managers* can be given tools to help identify, address and act on ethical questions that are faced in their work. There will be ethical trade-offs, and there will be risk management decisions to be made, which can be addressed only by the context-set stakeholders [4].

While ECCOLA has been developed, there has been a set of design principles applied. These same design principles need to apply to deployment model to form a coherent package of ECCOLA -tools. Those are considered in the following list. Additionally, several highlights from the AI ethics -research field have been raised to be considered in the conceptual framework creation.

The applied design principles (can be considered as requirements) are: (1) flexible and easy to use (ECCOLA), (2) modular to fit a wide variety of contexts (ECCOLA), (3) suitable for agile development (ECCOLA), (4) iterative, a cyclic process rather than linear [26,4], (5) actionable (ECCOLA, e.g. [2&27]), (6) non-bureaucratic (ECCOLA), (7) motivating and encouraging rather than judging (ECCOLA), (8) multi-disciplinary [4, 27], (9) facilitating participation and stakeholder involvement ([4,26,25]), (10) supporting systematic ethical impact analysis when ethical trade-offs arise ([4,26]), (11) applicable through and at any point in software product's lifecycle ([4,27]).

As the scope of this study is limited to a conceptual model, no evidence can be presented without further testing how the framework adheres to all the listed items that its design is based on.

*B. Deployment process*

A set of processes forming the deployment model is presented. ECCOLA is designed to be non-linear, constructive and modular instead of an iteratively repeated linear process. The software developers applying it deemed it as "(…) just another process tackled onto their other processes" [3]. The presented framework does not interfere with ECCOLA's core practices, for example, with introducing an iterative process, but it extends around it for several purposes. Those purposes are listed in the design principles in the previous section. Two quotes from Peters et al [4] aptly stress several of the design choices:

> *It is impossible to provide ethical principles that will be specific enough to give answers in practice yet broad enough to apply universally. But it is possible to provide a process. While every team and organization may devise their solutions to ethical dilemmas, they should have systematic processes to do so consciously and rigorously, leaving a record of these processes and the values and rationale they employed to make decisions. Such a process will not guarantee a product has no negative consequences, but it will help mitigate the risks and provide professionals with the reassurance of acting responsibly. (…)*

*Ethical implications should be considered early on and throughout the design, development, and implementation phases since value-laden trade-offs are often made even during the earliest stages of design. Therefore, ethical impact evaluation must be an ongoing, iterative process—one that involves various stakeholders at every step, and can be re-evaluated over time, and as new issues emerge.*

Fig. 1 presents the deployment end-to-end process that is proposed to be used to extend the existing ECCOLA method. The items in the model are subject to be adapted to differing needs and thus are provided in most cases only by naming the sub-processes and leaving the interpretation and implementation for the practitioner, i.e., avoiding making the framework too rigid or suitable only for a limited set of contexts.

The sub-processes (1.1 through 4.5) are elaborated as follows:

1.1 Any actor having a vested interest in the AI product should be involved in the entire framework and most desirably starting from the very beginning of the process, not merely the usual software development-centric roles. This is due to the fact that many ethical questions involve expertise the developing roles do not necessarily have, e.g., compliance, corporate risk management

1.2 Facilitating communication, participation, and shared vision creation of what the product will be once released and ensuring that ECCOLA method and this framework extension are established among stakeholders

1.3 A central process for the whole model: each stakeholder is presented with 21 tokens. They are to distribute the 21 tokens between ECCOLA cards as per their understanding of what cards are more relevant than others when considering the whole product once complete and in use.

The previous notion also reflects the idea that trade-offs are needed in most cases, and not all ethical considerations can be considered equally [4]. As contexts and motivations vary, it is necessary to understand what things matter the most in this context and which items perhaps have no impact at all. The card with the highest number of assigned tokens (sum of tokens given by each stakeholder) is the highest priority.

The one with the least number of tokens (0 is acceptable) is the least priority. How these priorities translate to objectives and e.g., user stories or requirements is up to the team to plan in process 1.7

1.4 Potential triggers like upcoming requirements from an analysis of regulation impacting the product and its use, any stakeholder requiring the re-evaluation of prioritization and target state, and so on are identified and tracked in 2.4

1.5 Using the Situation picture chart introduced and explained in the next Section, an initial target state is formed to show how ECCOLA cards (i.e., ethical considerations) are situated in the chart concerning their relevance – this is the baseline to which outcomes are later compared plans versus outcomes

1.6 Involving and participating stakeholders on the prioritization results, what they mean for the product, and in terms of work to be done

1.7 A sprint plan or any plan identifying the set priorities and work or definition of work needed to realize those priorities is formed

2.1-2.3 these are the ECCOLA core processes [3]

2.4 A continuous sub-process governed by, e.g., the *product manager* or *design lead*. The process allows for identification of any previously agreed trigger to initiate the 3. -process

3.1 Each stakeholder can reflect the potentially changed environment of the developed product and reassign their prioritization tokens as they wish

3.2 Per outcome of 3.1 and other issues considered, the guiding instruments (e.g., requirements, user stories) of the development are adjusted accordingly, if needed

3.3 Prior plans are updated according to factors discovered in subprocess 3.1, as is the target state picture if changes are of sufficient impact

3.4 A process that is getting input from 3.1-3.3 and assuring that needed traceability of decision made and their ratification is documented and traceable

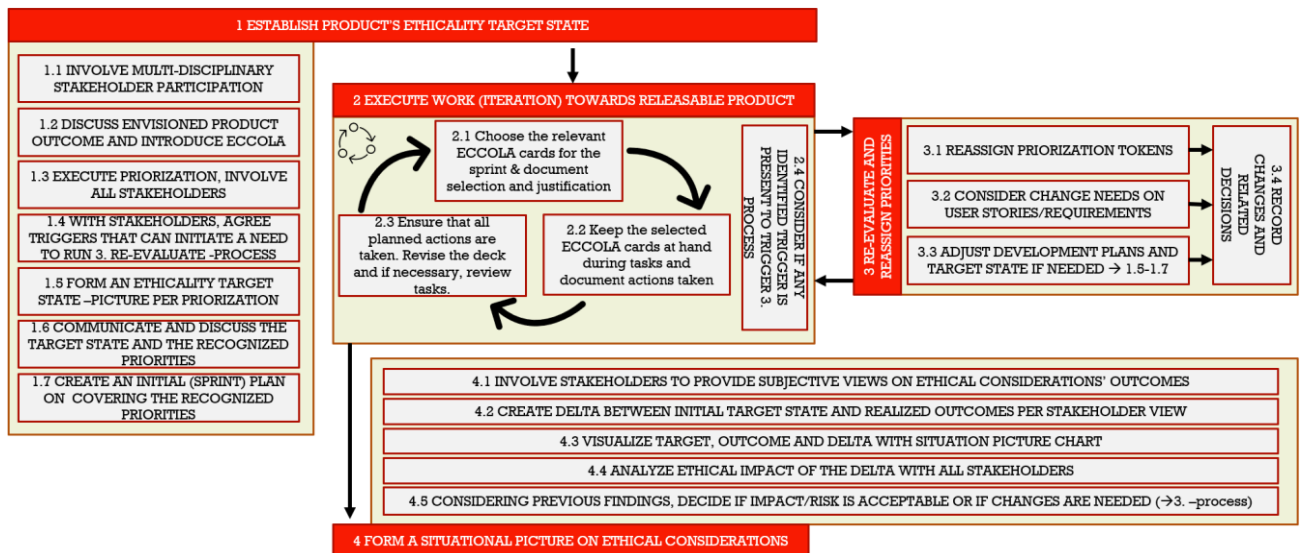

Fig. 1. The deployment process

4.1 As ethical factors are impossible to benchmark [3] (i.e., what is ethical enough) and highly subjective to context, metrics prove challenging to produce. Yet, some form of them is needed for communicating the ethicality of the product and the work involved. In this process, each stakeholder will score each ECCOLA card for her subjective assessment of how the card has been covered, considering its previously set priority. That is done with a simple 1-5 Likert-scale, for each card by each stakeholder: *1: strongly disagree, 2: disagree, 3: neither agree nor disagree, 4: agree and 5: agree strongly*. The average of the score is calculated and presented in 4.3

4.2 The latest outcomes from 3.1-3.3 are compared to the initial target state set in 1.5 to show how (and why) the target state has changed from the initial target as cards have been re-prioritized along the process

4.3 A visualization (Fig. 2) is presented to analyze, if the set target is met or not met and how the target has potentially changed during the product development process

4.4 Using the chart created in subprocess 4.3, needed sessions are facilitated involving all previous stakeholders to elaborate findings from 4.3 and if there is considerable delta, initiate needed ethical impact analyses, enterprise risk management process, or similar control method

4.5 Per outcome of subprocess 4.4, the created product is either "sufficiently ethical" or having flaws that need returning to the drawing board and back to process number 3.

*C. AI SW product's ethicality situational picture*

Within the previous sections, it has been noted that AI ethics are very subjective and sensitive to their considered context. Also, the current discussions on AI ethics are around ethical principles and if they can be created. Such principles have almost zero traction with what AI ethics practice is longing for – actionable tools. How can ethical considerations be managed, evaluated, and communicated in such way that all stakeholders can understand the communication?

As part of the presented framework and in addressing the needs of such a communication, participation, and planning-aid tool, an effort to contribute is presented. Respecting the principles (requirements) set in section A, an ethical situational picture Fig. 2 (visualization) is created. An example of how the results emerging from using the model with ECCOLA is shown in Fig. 2. The numbers correspond to ECCOLA card numbers [3], e.g., #8 is *data quality*.

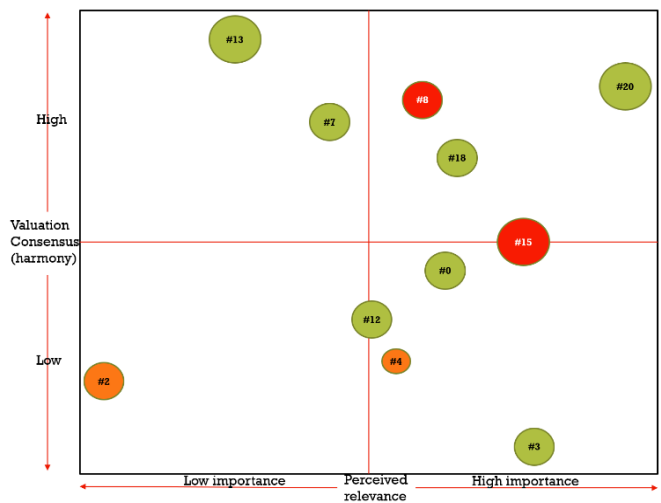

Fig. 2. An example of assessment results from an imaginary AI product

The situational picture (Fig. 2) is populated according to the following rules:

- the perceived relevance-axis scale is created so that the ECCOLA card receiving the lowest token amount defines "low importance" while the card receiving the highest token amount defines the "high importance" -boundary. The middle of the axis is determined by the card coming closest to the average token amount of the cards

- valuation consensus axis communicates how an agreement on the relevance of each ECCOLA card the stakeholders has been arrived upon. Thus, the card with the highest valuation defines the upper boundary of the axis, and the one with the lowest consensus defines the bottom perimeter. The card that comes closest to the consensus difference score represents the axis centre

- valuation consensus score is defined for each card by identifying the median of the token assignments from each stakeholder and then analyzing how many of the assigned priorities differ more than one token over or under from the median value. The number of such deviations is the (negative) harmony score. The card with the lowest harmony score has the highest harmony among stakeholders' views on priorities. In contrast, the one with the highest score communicates that the stakeholders' opinions on that card are most mixed

- bubble color communicates the outcome of sub-process 4.1, i.e., the stakeholders' view on how well each card has been covered in the development work done. The value is created by calculating the average of Likert scores assigned by the stakeholders in subprocess 4.1. In the range of 4...5, the bubble is green. 3…4 is indicated by yellow, and values below three are designated as red

- size of the bubble (small, medium, large) conveys if the card's priority has been changed along the software development process. No changes correspond to default medium size. An increase in priority is indicated by a large bubble and a decrease in priority by a small bubble

- presented bubble size -coding can be replaced with "initial state"-instance of the card presented in the chart, e.g., by a gray bubble with a connector between the initial relevance score position and the position at subprocess 4.3

The presented situational picture (Fig. 2) is an integral part of the model and works in unison with the deployment process. It starts from the beginning of drafting out the target state of the product's ethical considerations and priorities and ending up through the iterative processes in between to convey to the stakeholders (e.g., management) the ethicality of the product at hand. Although very simple, it might turn out to be a valuable communication tool and evidence of ethical decisions made during the product lifecycle.

Use cases for the ethicality situational picture should be many according to the first author's experience. The need of communication and involvement are paramount, and the concept of the situational picture presented is created with that need in mind, not forgetting the simple use of the tool.

## IV. Discussion and conclusions

Applying the design science research methodology (DSRM), a conceptual deployment model to extend ECCOLA was created. Apart from the context of AI SW development, the framework is also applicable to the use-phase of such AI products and governing and operating them in different management disciplines such as ICT service management [28], where applications are constantly evolving and developed further.

The study is deemed valuable for both researchers and practitioners seeking to have an actionable AI ethics-enabling tool for application to the AI product context. The model facilitates the adoption of ECCOLA. In addition, it offers a light ethicality assessment capability for AI SW development as well as for operational use-phase. Further, it provides practical insight into the application possibilities of ECCOLA.

As the model is still a conceptual artefact, additional study is needed to satisfy the demonstration and evaluation needs via applying the model in multiple case studies. Other research opportunities would describe how the framework supports the post-development lifecycle of an AI SW product and what other use cases, like standardizing the evaluation scoring for different categories of AI products, would be feasible. One interesting opportunity is integrating enterprise risk management (RM) practices and processes with the model and thus potentially acquire traction and interest from e.g., senior management or the institution / corporate board.